%% file: QCD-10-013_temp.tex
\pdfoutput=1

\documentclass[11pt,twoside,a4paper,cmspaper,final,collab]{cms-tdr}
\def\svnVersion{35568:35574}\def\cmsCernNo{2010-072}\def\cmsCernDate{\today}\def\cmsMessage{Submitted to Physics Letters B}

\begin{document}\cmsNoteHeader{QCD-10-013}

\hyphenation{had-ron-i-za-tion}
\hyphenation{cal-or-i-me-ter}
\hyphenation{de-vices}
\RCS$Revision: 35574 $
\RCS$HeadURL: svn+ssh://alverson@svn.cern.ch/reps/tdr2/papers/QCD-10-013/branches/plb-0/QCD-10-013.tex $
\RCS$Id: QCD-10-013.tex 35574 2011-01-31 19:17:04Z alverson $
\input{ptdr-definitions}
\cmsNoteHeader{QCD-10-013} 
\title{First Measurement of Hadronic Event Shapes in pp Collisions at $\sqrt{s}=7$\TeV}

\ifthenelse{\boolean{cms@external}}{}{}

\newcommand{\Red}{\color{red}}

\newcommand{\lumifinal}{{3.2\,pb${}^{-1}$}}
\newcommand{\pythiasix}{\PYTHIA{6}}
\newcommand{\herwigpp}{\HERWIG{++}}
\newcommand{\pythiaeight}{\PYTHIA{8}}
\newcommand{\generators}{\pythiasix, \pythiaeight, \herwigpp, \MADGRAPH, and \ALPGEN}
\newcommand{\ptone}{\ensuremath{\pt{}_{,1}}}
\newcommand{\lowbin}{\ensuremath{90 < \ptone < 125\GeVc}}
\newcommand{\medbin}{\ensuremath{125 < \ptone < 200\GeVc}}
\newcommand{\highbin}{\ensuremath{\ptone > 200\GeVc}}

\date{\today}

\abstract{
Hadronic event shapes have been measured in proton-proton collisions at $\sqrt{s}=7\,$TeV,
with a data sample collected with the CMS detector at the LHC. The sample corresponds to an integrated luminosity of~3.2\,pb${}^{-1}$.
Event-shape distributions, corrected for detector response, are compared with five models of QCD multijet production.}

\hypersetup{%
pdfauthor={CMS Collaboration},%
pdftitle={First Measurement of Hadronic Event Shapes in pp Collisions at sqrt(s)=7 TeV},%
pdfsubject={CMS},%
pdfkeywords={CMS, LHC, QCD, event shapes}}

\maketitle

Event shapes provide information about the properties of hadronic final states from particle collisions.
Suitably defined event-shape variables were among the first observables proposed to test the theory of quantum chromodynamics
(QCD)~\cite{Biebel:2001dm,Kluth:2006bw} and have been important in enabling progress in the theory.
At $\mathrm{e^{+}e^{-}}$ and $\mathrm{ep}$ colliders, event shapes have played a crucial role in the
extraction of the strong coupling constant $\alpha_{s}$.
They have been essential in tuning the parton shower and non-perturbative components
of Monte Carlo (MC) event generators and have provided a laboratory for developing and testing
analytical probes of the hadronization process.
More recently, a large set of event-shape variables suitable for pp colliders
has been proposed~\cite{Banfi:2010xy}. An important aspect of these variables
is their normalization to the measured sum of transverse momentum or energy of all the objects in the event.
It is thus expected that energy-scale uncertainties should cancel to a large extent.
Event-shape variables represent a valuable tool for early measurements of the properties of QCD multijet events
at the Large Hadron Collider (LHC) and the tuning of MC models~\cite{Wraight:2011ej}.

This Letter presents the first measurement of hadronic event shapes with a data sample of
7\TeV proton-proton collisions collected with the Compact Muon Solenoid (CMS)
detector at the LHC. The data sample corresponds to an integrated luminosity of \lumifinal.

A detailed description of the CMS experiment can be found elsewhere~\cite{:2008zzk}.
CMS uses a right-handed coordinate system, with the origin located at the
nominal collision point, the $x$-axis pointing towards the center of the LHC ring, the $y$-axis pointing
up (perpendicular to the LHC plane), and the $z$-axis along the anticlockwise beam direction.
The polar angle $\theta$ is measured from the positive $z$-axis, the azimuthal angle $\phi$
is measured in the $xy$ plane, and the pseudorapidity is defined as $\eta=-\ln[\tan(\theta/2)]$.
The central feature of the CMS apparatus is a superconducting solenoid, of 6\,m
internal diameter, providing an axial field of 3.8\,T. Within the field volume are the silicon pixel and strip tracker,
the crystal electromagnetic calorimeter (ECAL), and the brass/scintil\-lator hadron calorimeter (HCAL).
Muons are measured in gas-ionization detectors embedded in the steel return yoke.
In the region $\vert \eta \vert< 1.74$, the HCAL cells have widths of 0.087 in pseudorapidity and 0.087\,rad in azimuth ($\phi$).
In the $(\eta,\phi)$ plane, and for $\vert \eta \vert< 1.48$, the HCAL cells map on to $5 \times 5$ ECAL crystal arrays
to form calorimeter towers projecting radially outwards from close to the nominal interaction point.
At larger values of $\vert \eta \vert$, the size of the towers increases and the matching ECAL arrays contain fewer crystals.
A preshower detector consisting of two planes of silicon sensors interleaved with lead is located in front
of the ECAL at $|\eta|>1.479$.
In addition to the barrel and endcap detectors, CMS has extensive forward calorimetry covering the region $3.0<|\eta|<5.0$.

Two event-shape variables have been studied: the central transverse thrust $\tau_{\perp,\mathcal{C}}$ and
the central thrust minor $T_{m,\mathcal{C}}$. The two variables probe different QCD radiative processes
and are mostly sensitive to the modeling of two- and three-jet topologies. The term {\em central}
($\mathcal{C}$) indicates that the input to the calculation of these quantities are jets in the central region
of the detector ($|\eta|<1.3$), where sub-leading contributions in the calculation of the event-shape variables
are less significant, and systematic uncertainties on the jet reconstruction are smaller.

The central transverse thrust is defined as~\cite{Banfi:2010xy}
\begin{equation}\label{eq:centrthr}
\tau_{\perp,\mathcal{C}}\equiv 1-\max_{\hat{n}_{\text{T}}}\frac{\sum_{i}\left|\vec{p}_{\perp,i}\cdot \hat{n}_{\text{T}} \right|}{\sum_{i}p_{\perp,i}},
\end{equation}
where $p_{\perp,i}$ is the transverse momentum of selected jet $i$.
The axis $\hat{n}_{\text{T}}$ which maximizes the sum, and thus minimizes $\tau_{\perp,\mathcal{C}}$,
is called the thrust axis $\hat{n}_{\text{T},\mathcal{C}}$.
The central transverse thrust is a measure of the momentum in the plane defined by $\hat{n}_{\text{T},\mathcal{C}}$
and the beam axis. The central thrust minor is a measure of the momentum out of this plane and is defined as
\begin{equation}\label{eq:centhrmin}
T_{m,\mathcal{C}}\equiv  \frac{\sum_{i}\left|\vec{p}_{\perp,i}\times\hat{n}_{\text{T},\mathcal{C}}\right|}{\sum_{i} p_{\perp,i}}.
\end{equation}
Two-jet events that are well balanced have low values of these two variables, while isotropic multijet events have high values.

The transverse momenta of jets are used as input to the event-shape calculation.
Jets are reconstructed using individual particles that have been identified, and whose energies have been measured,
using a particle flow technique~\cite{CMS-PAS-PFT-10-002}, which combines information from all subdetectors:
charged tracks in the tracker and energy deposits in the electromagnetic and hadronic calorimeters, as well as signals in
the preshower detector and the muon system. The energy calibration is performed separately for each particle type.
As a result, the input to the jet clustering is almost fully calibrated and the resulting jets require only a
small energy correction (below 10\% in the central region). Jet clustering is performed using the anti-$k_{T}$ clustering
algorithm~\cite{Cacciari:2008gp} with a distance parameter $R=0.5$.

Five MC generators are used to produce simulated samples for comparison with the data; the specifics of each generator are detailed below.
In addition to the generator-level samples, we use ``full-simulation'' samples, where the events produced
at the generator level are processed with a simulation of the CMS detector response based on {\sc Geant4}~\cite{Agostinelli:2002hh}.
As event-shape distributions are sensitive to
QCD radiation, they are primarily affected by the description of the parton showering and the hadronization
process, and, to a lesser extent, by the description of multiparton interactions,
which is included in all generators used.

The first generator considered is \PYTHIA6.4.22 (\pythiasix)~\cite{Sjostrand:2006za} with tune D6T~\cite{Field:2010su}.
In this version of \PYTHIA, parton showers are ordered by mass.
The second generator is \PYTHIA8.145 (\pythiaeight)~\cite{Sjostrand:2007gs} with tune 2C~\cite{Corke:2010fk}.
In this version of \PYTHIA, parton showers are ordered by $\pt$.
The underlying event model is based on the multiple-parton interaction model of \pythiasix,
interleaved with initial- and final-state radiation.
The third generator is \herwigpp~2.4.2~\cite{Bahr:2008pv} used with the tune of older version~2.3.
The parton showering in \herwigpp\ is based on the coherent-branching algorithm, with angular ordering of the showers.
The underlying event is simulated using an eikonal multiple parton-parton scattering model.
The fourth is \MADGRAPH~4.4.24~\cite{Alwall:2007st} in conjunction with \pythiasix, with tune D6T.
Events containing from two to four jets matched to partons with \pt\ above $20\GeVc$ are produced with \MADGRAPH
using a matrix element (ME) calculation and subsequently passed to \PYTHIA to generate parton showers (PS).
The MLM matching procedure~\cite{Mrenna:2003if} is used to avoid double counting between the ME and
PS calculations. For the matching, the minimum jet $\pt$ threshold is set to $30\GeVc$.
Finally, the \ALPGEN~2.13~\cite{Mangano:2002ea} generator is used in a similar way to \MADGRAPH.
\ALPGEN samples are produced separately
for each jet multiplicity from two to six jets, matched to partons with \pt above $20\GeVc$, and are weighted  according to their theoretical cross section.
Events produced with \ALPGEN using the ME calculation are passed to \PYTHIA, and the MLM matching procedure is used to avoid double counting.
For the matching of ME partons to jets, the lower jet $\pt$ threshold is set to $20\GeVc$ and the maximum distance between
partons and jets is kept to its default value of $\Delta{}R=0.7$.

The data were collected between April and August 2010.
Noncollision background is removed by applying quality cuts that ensure the presence of a well-reconstructed primary vertex~\cite{CMS-PAS-TRK-10-005}.
The selected data sample is then divided into three bins defined by $\ptone$, the \pt\ of the leading jet (the jet reconstructed offline with the highest \pt).
The low-\pt bin contains events with \lowbin, the medium-\pt bin with \medbin, and the high-\pt bin
with \highbin. Data in all bins are selected using single-jet triggers that require an online reconstructed
jet with \pt greater than $30\GeVc$ for the low-\pt bin, and greater than $50\GeVc$ for the medium-\pt\ and
high-\pt bins. The trigger with a $30\GeVc$ threshold was prescaled as the instantaneous luminosity of
the LHC increased; the effective luminosity in the low-\pt\ bin is only 0.32\,\pbinv.
The trigger efficiency, measured from a sample acquired with lower-threshold triggers,
is greater than $99\%$ for all \pt bins.

Quality cuts are imposed on the jets in order to remove spurious jets caused by calorimeter noise or other remaining noncollision background.
Jets must consist of at least two particles, including at least one charged hadron, and not more than 99\% of the jet energy
may be carried by neutral hadrons alone, by photons alone, or by electrons alone.
All jets within the detector acceptance ($|\eta|<5)$, with $\pt> 30\GeVc$,
and passing the quality criteria are subsequently considered.
However, if one of the two leading jets does not pass the quality cuts, the event is rejected.
This requirement rejects less than 1\% of the events.
After the initial jet selection, the two leading jets are required to be within $|\eta|<1.3$, or the event is rejected.
All selected jets within $|\eta|<1.3$ are used in the event-shape calculation.

In the low, medium, and high-\pt data samples, this selection retains respectively
$62\,000$, $180\,000$, and $23\,000$ events, of which 77\%, 65\%, and 52\% are events with
exactly two selected jets.

The event-shape distributions are distorted by the energy and angular resolutions of the detector.
The measured distributions are unfolded to allow comparison
with the event-shape distributions calculated in the generator-level samples.
We use a regularized unfolding method based on singular-value decomposition (SVD) of the response matrix~\cite{Hocker:1995kb}.
The inputs to the unfolding algorithm are the distributions measured in data
and the response matrix, determined from the event-shape distributions
of the \pythiasix\ generator-level and full simulation samples.
The algorithm returns the unfolded data distribution, together with its correlation and error matrices.
All of the unfolding corrections are below 5\%.

The dominant systematic uncertainty in the event-shape distributions comes from the jet energy scale.
While the event-shape definitions are expected to be invariant under a shift in jet energy scale, the uncertainty
on the energy scale modifies the number of jets passing the \pt threshold, thus affecting the event shapes.
In order to estimate the resulting uncertainty on the event-shape distributions, a shift of the jet energy scale is applied
to all jets entering the calculation, based on an uncertainty estimate which varies between~3 and~5\%, depending on $\eta$ and \pt~\cite{CMS-PAS-JME-10-010}.
The maximum bin-by-bin difference between the original distributions and the two shifted distributions,
after unfolding, is of the order of 4\% and is assigned as a systematic uncertainty.
The effects of the angular resolution and the uncertainty on the jet energy are estimated in a similar way and are found to be insignificant.

Energy resolution studies have revealed up to a 10\% difference between the data and the MC simulation~\cite{CMS-PAS-JME-10-010}
from which the response matrix is estimated. In order to estimate the effect of this difference on
the event-shape distributions, a new response matrix is determined from a MC sample in which the jet energies
are smeared by an additional 10\%. The differences between the original event-shape and the unfolded event-shape distributions
obtained with the new response matrix are found to be below 1\%.

The robustness of the result against possible bias due to the unfolding was also tested.
We checked that the differences between the unfolded data and each generator-level sample
were comparable with the differences found between data before unfolding and the corresponding full-simulation sample.
Here, the agreement between event-shape distributions in data and simulation was quantified
by a $\chi^{2}$ statistic.
We also ranked the simulated samples by their increasing agreement with the data, and checked that the ranking
was the same before and after unfolding.
Alternative unfolding methods~\cite{D'Agostini:1994zf} were applied; the resulting event-shape distributions differed by less
than 1\% from the results of the SVD unfolding. We also determined the response matrix using \MADGRAPH MC samples
instead of \pythiasix\ samples and found no significant differences.

The increase in the instantaneous luminosity of the LHC was accompanied by an increase in the average number of interactions per bunch crossing.
The effect on the event-shape distributions was checked by measuring these distributions
separately for events with exactly one, two, and more than two primary vertices. The resulting three sets of
distributions were found to agree within statistical errors.

Finally, we constructed event-shape distributions with an alternative jet reconstruction, based exclusively on information
from the tracking detectors~\cite{CMS-PAS-JME-10-006}, and found that they agree within errors with the
distributions based on particle flow reconstruction.

\begin{figure*}[hbtp]
  \begin{center}
    \includegraphics[width=0.49\textwidth]{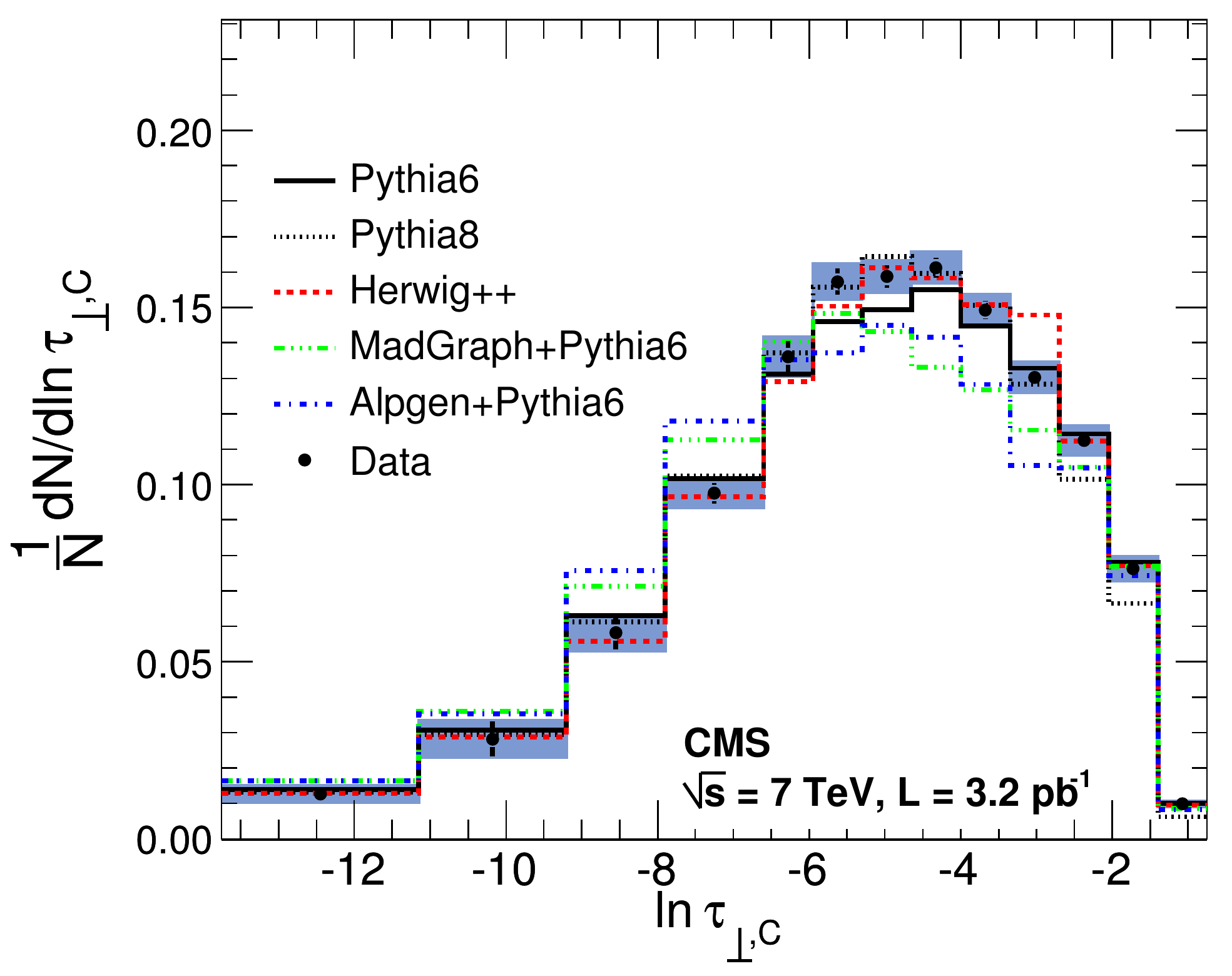}
    \includegraphics[width=0.49\textwidth]{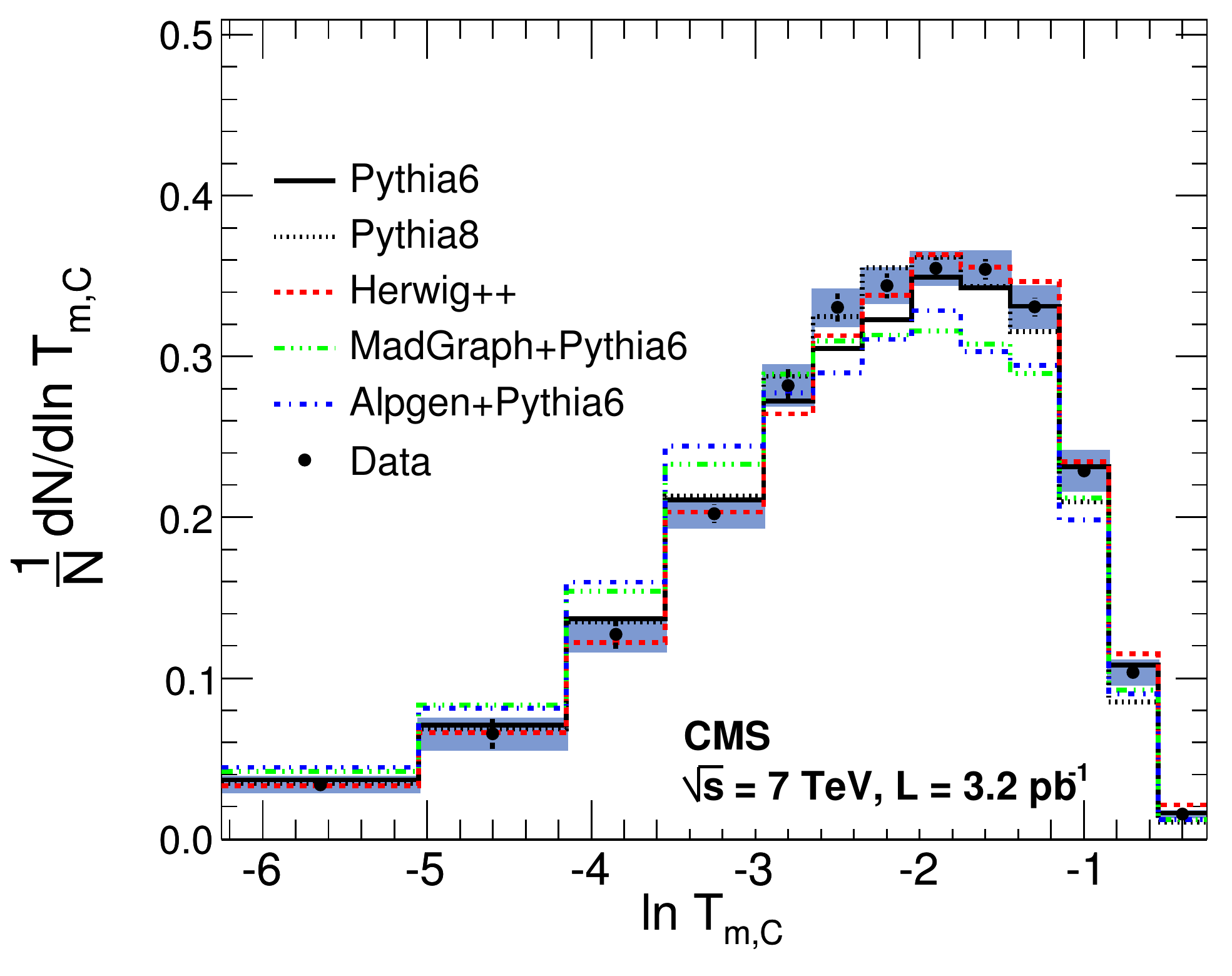}
    \includegraphics[width=0.49\textwidth]{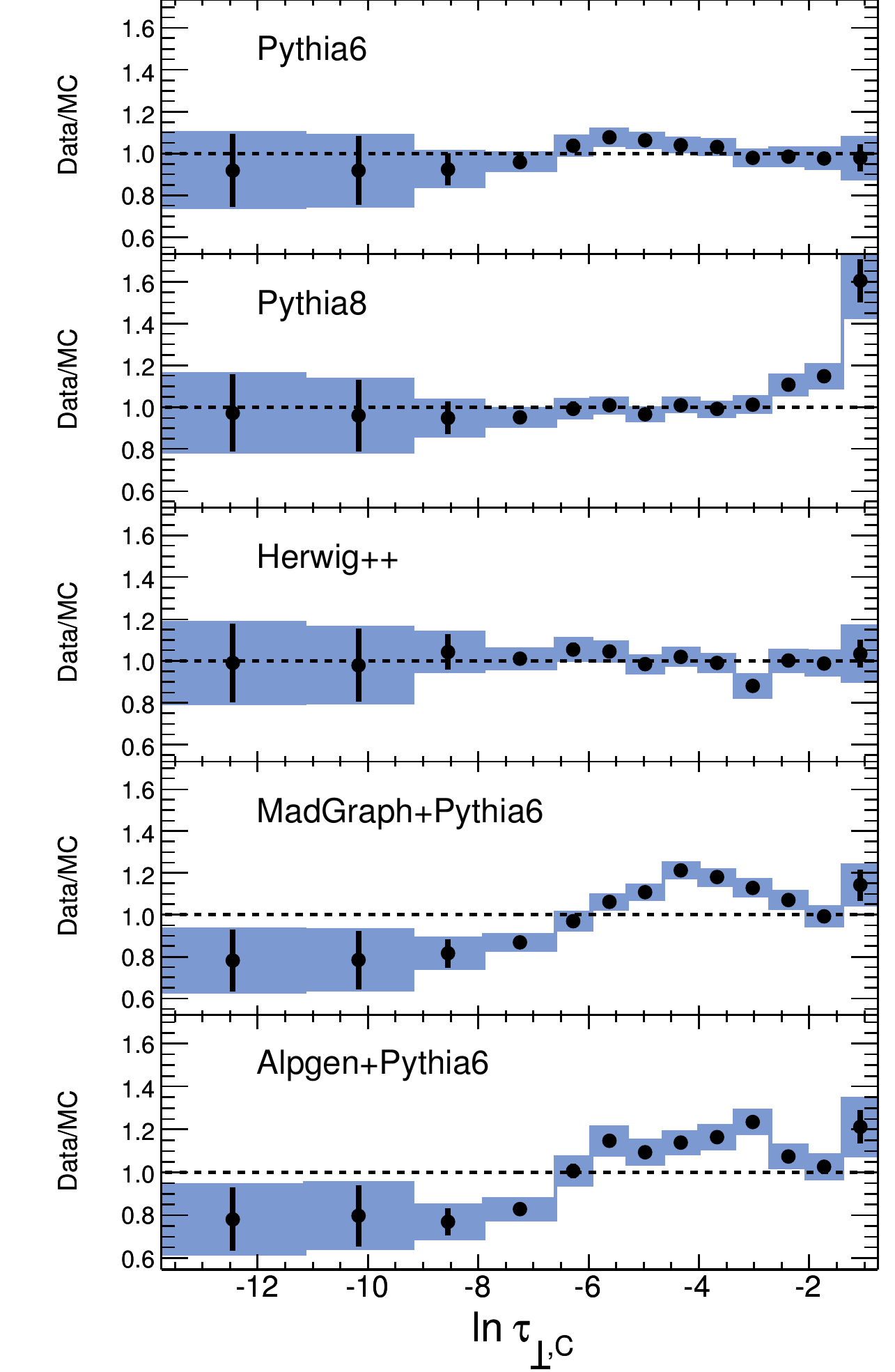}
    \includegraphics[width=0.49\textwidth]{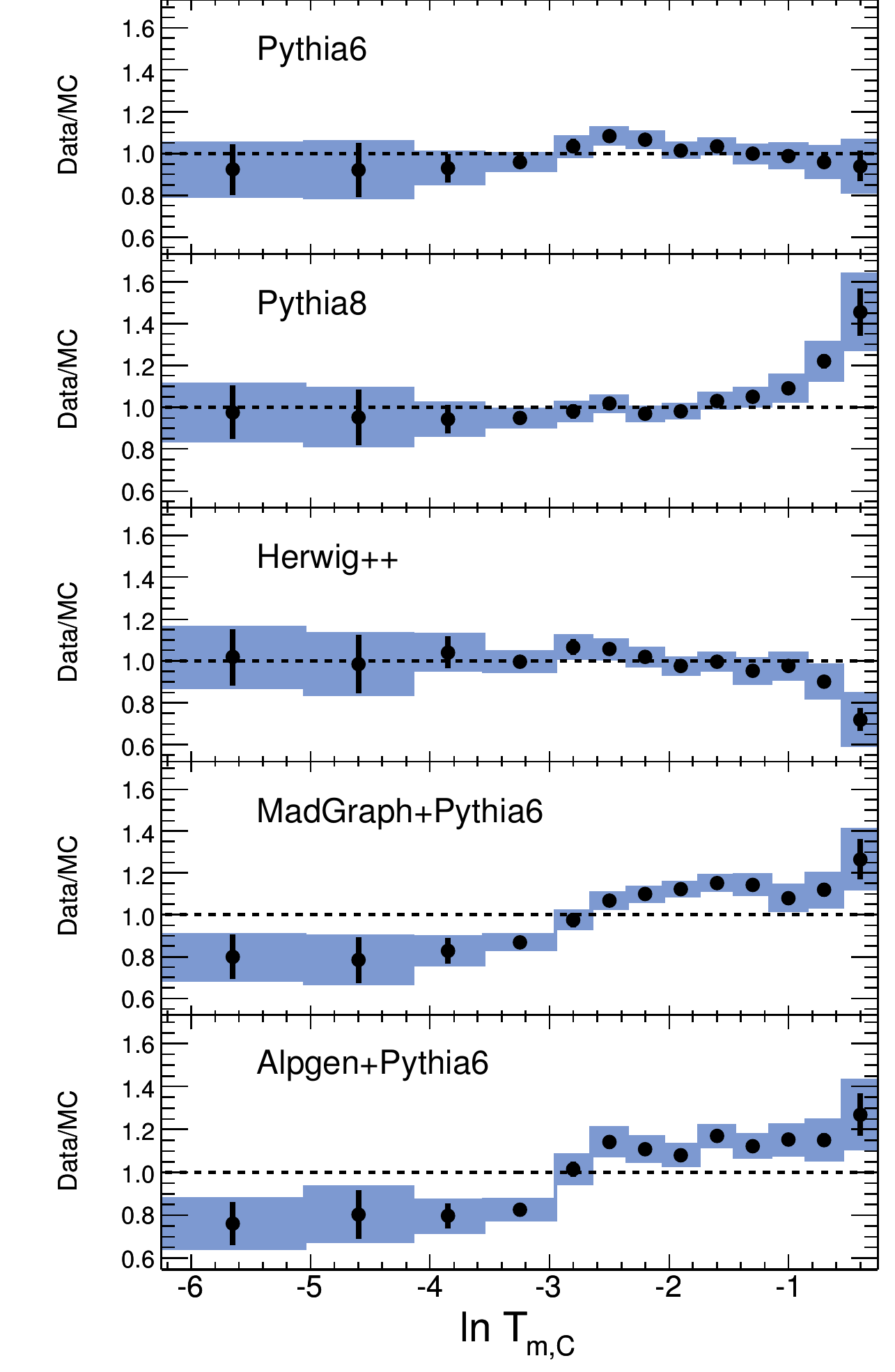}
    \caption{Distributions of the logarithm of the central transverse thrust (top left) and central thrust minor (top right)
             for events with a leading jet \pt\ between 90 and 125\GeVc, from data and from five MC simulations.
    				   The error bars on the data points represent  the statistical uncertainty on the data, and
				     the shaded (blue) bands represent the sum of statistical and systematic errors.
             The lower plots show the ratio between data and the different simulated samples.}
    \label{fig:comparison1}
  \end{center}
\end{figure*}

\begin{figure*}[hbtp]
  \begin{center}
    \includegraphics[width=0.49\textwidth]{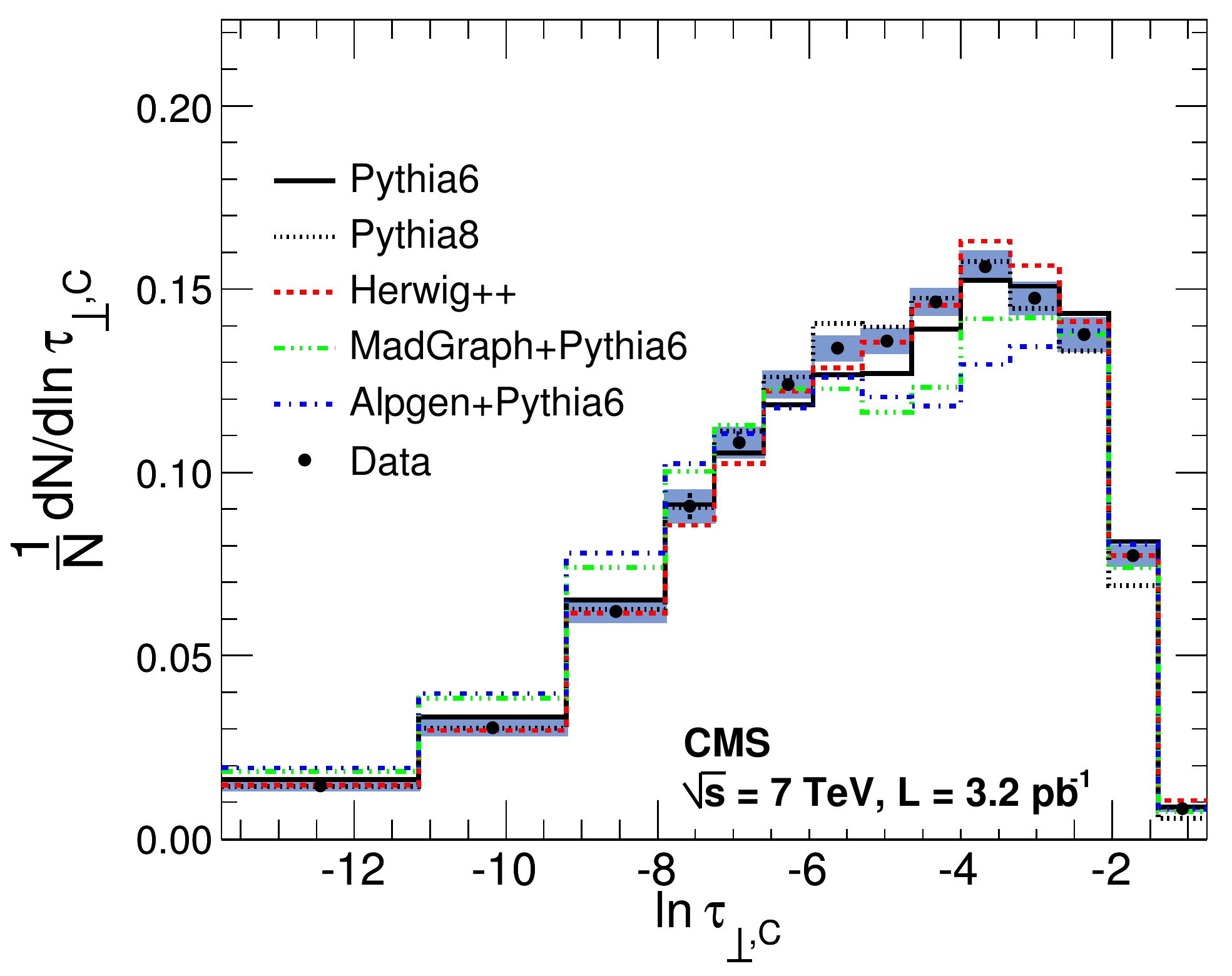}
    \includegraphics[width=0.49\textwidth]{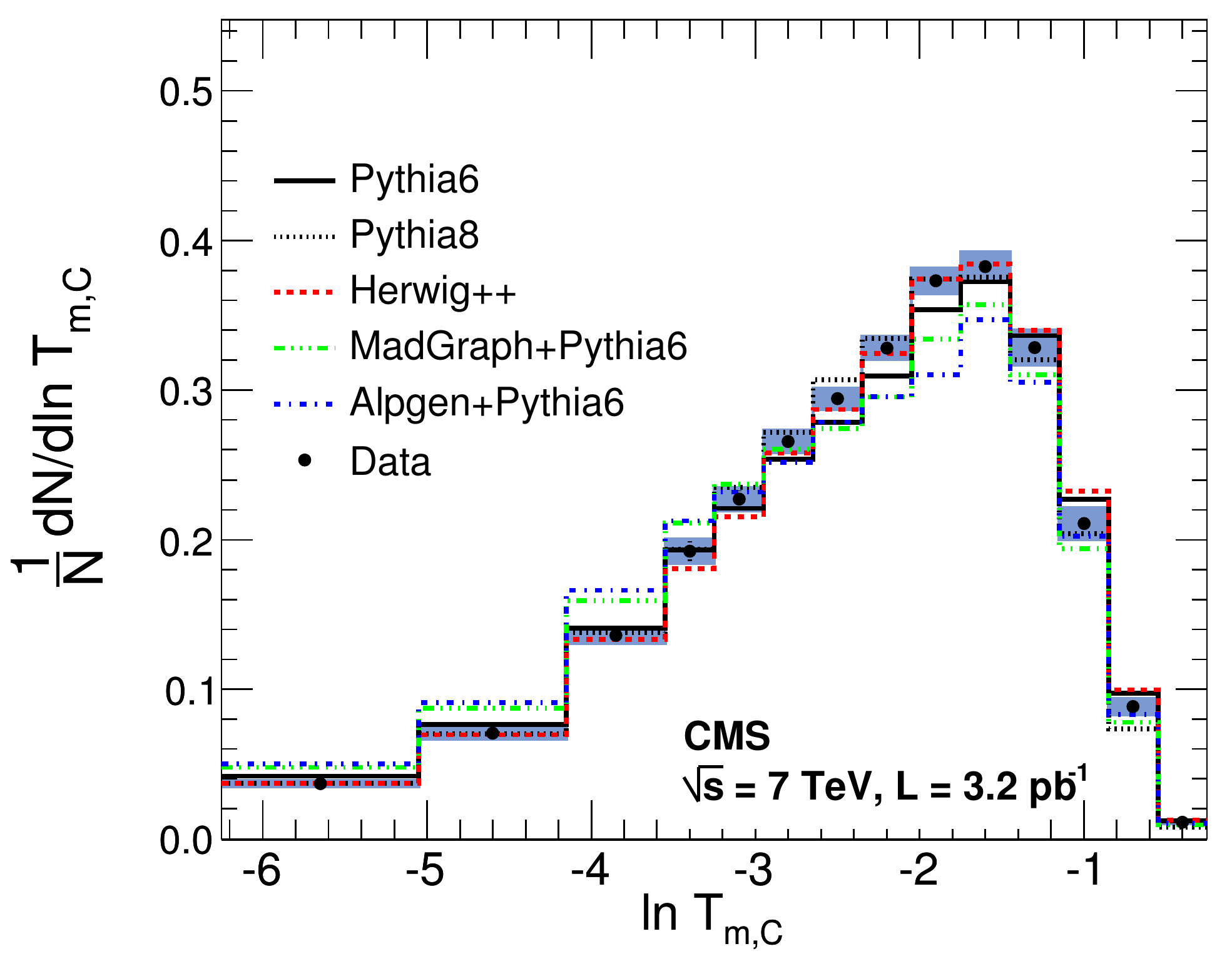}
    \includegraphics[width=0.49\textwidth]{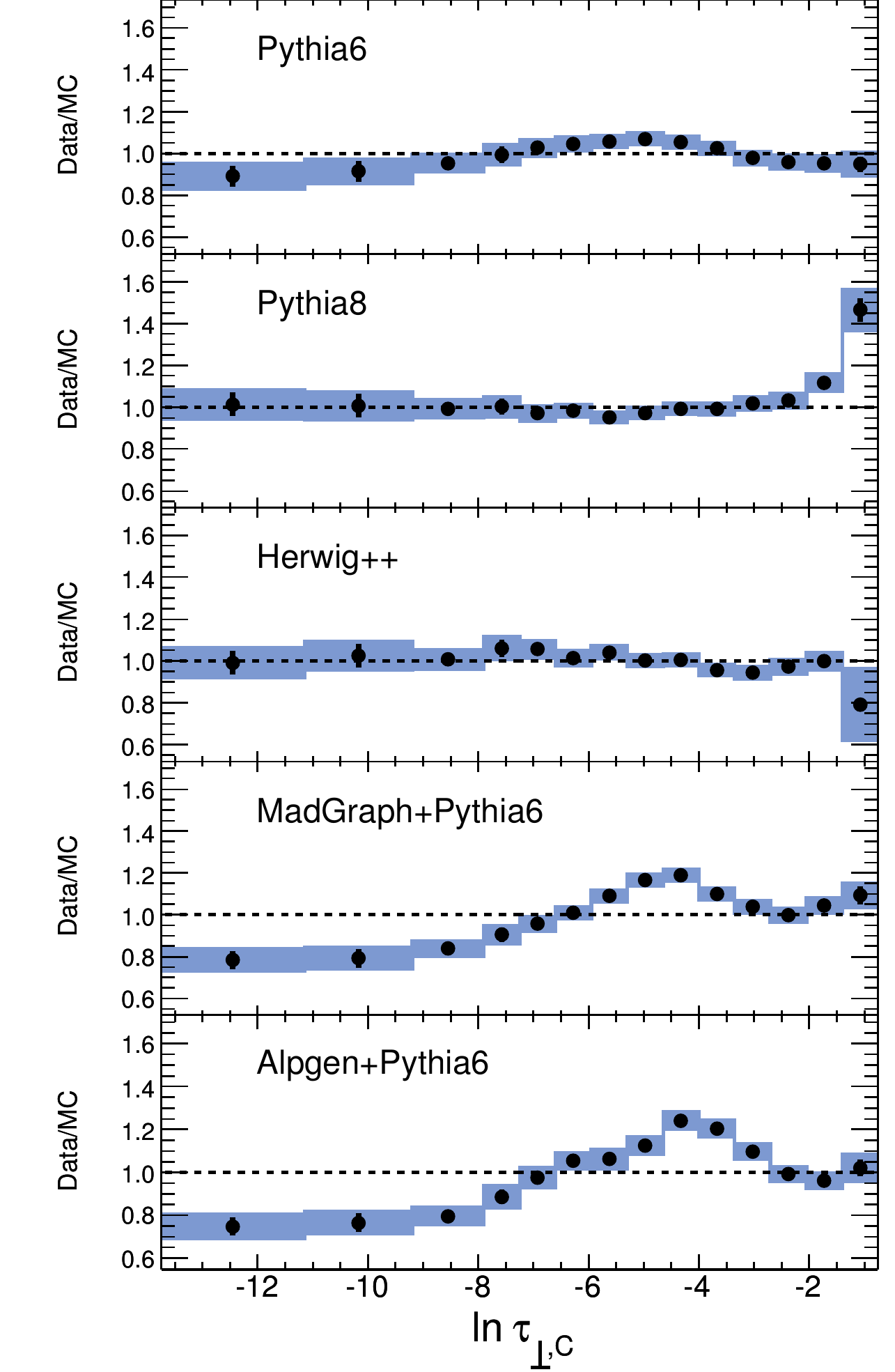}
    \includegraphics[width=0.49\textwidth]{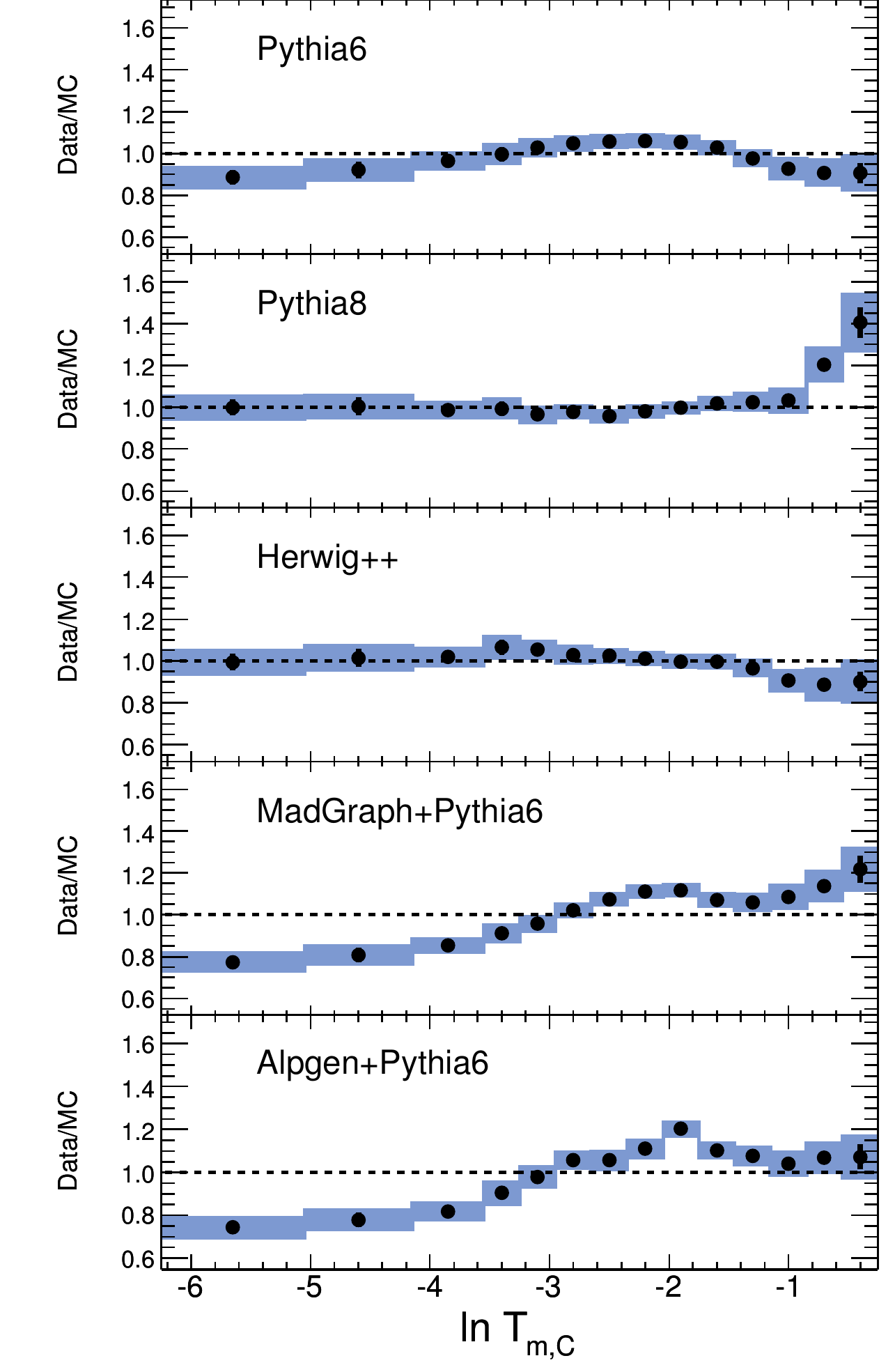}
    \caption{Distributions of the logarithm of the central transverse thrust (top left) and central thrust minor (top right)
             for events with a leading jet \pt\ between 125 and 200\GeVc, from data and from five MC simulations.
    				   The error bars on the data points represent  the statistical uncertainty on the data, and
				     the shaded (blue) bands represent the sum of statistical and systematic errors.
             The lower plots show the ratio between data and the different simulated samples.}
    \label{fig:comparison2}
  \end{center}
\end{figure*}

\begin{figure*}[hbtp]
  \begin{center}
    \includegraphics[width=0.49\textwidth]{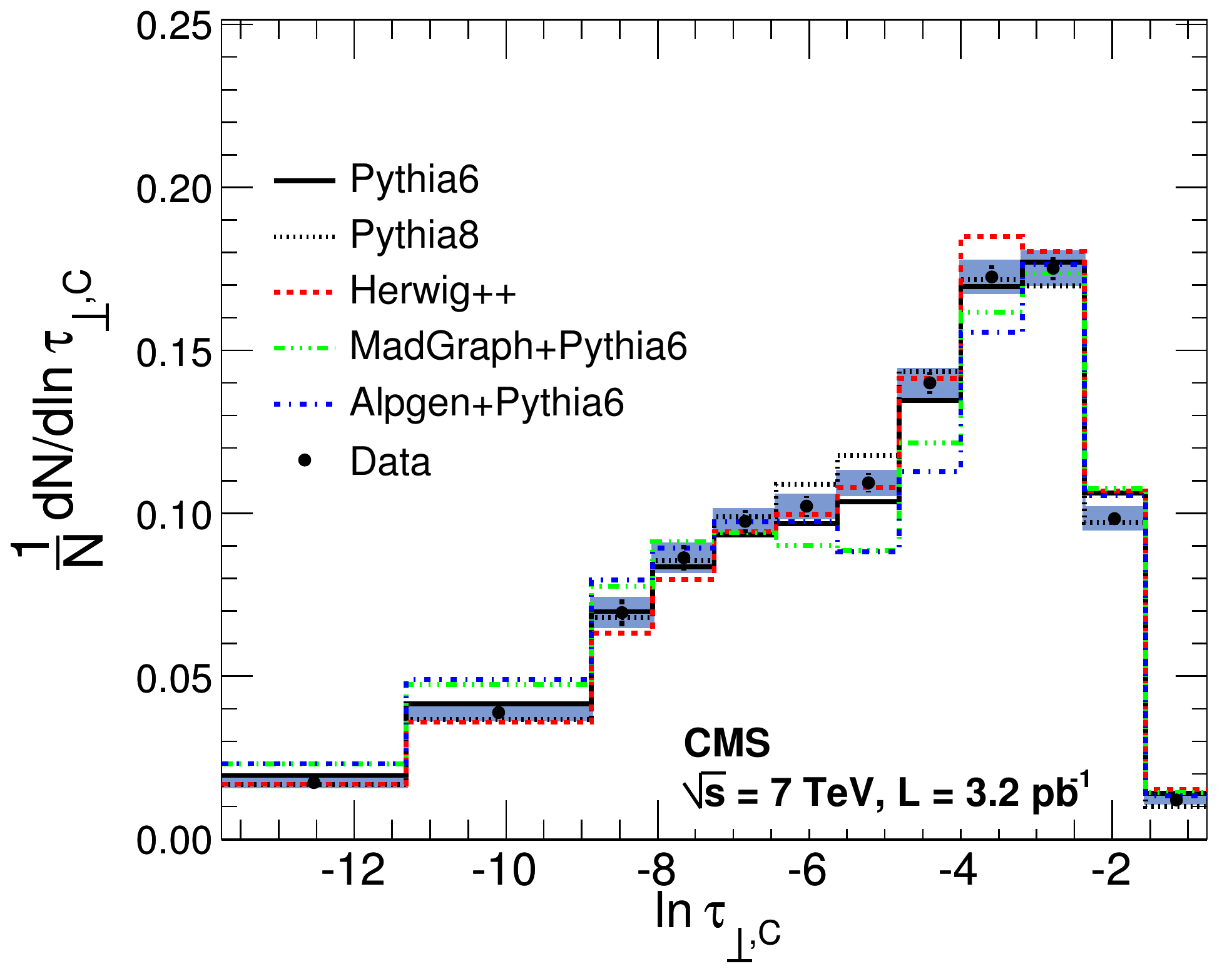}
    \includegraphics[width=0.49\textwidth]{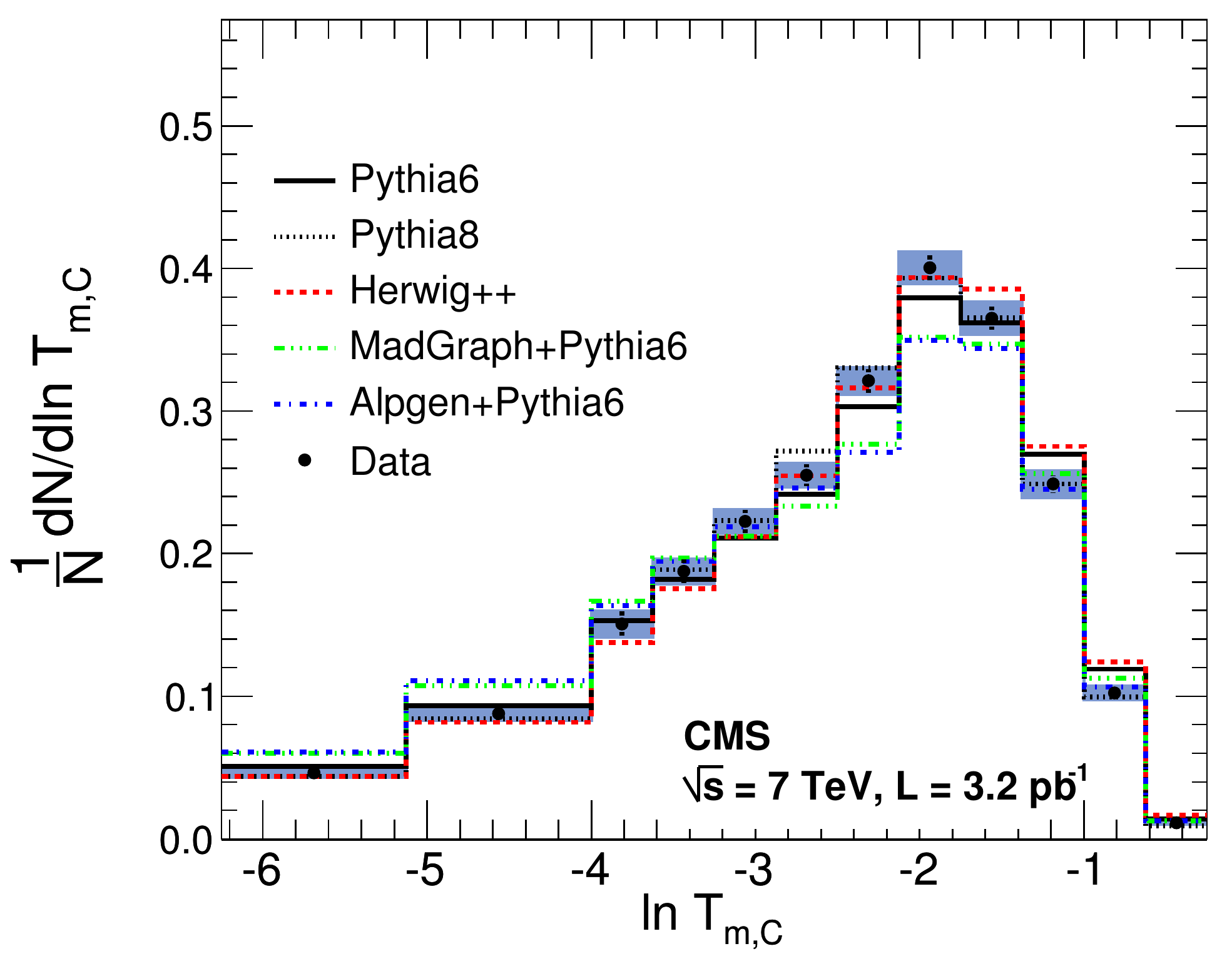}
    \includegraphics[width=0.49\textwidth]{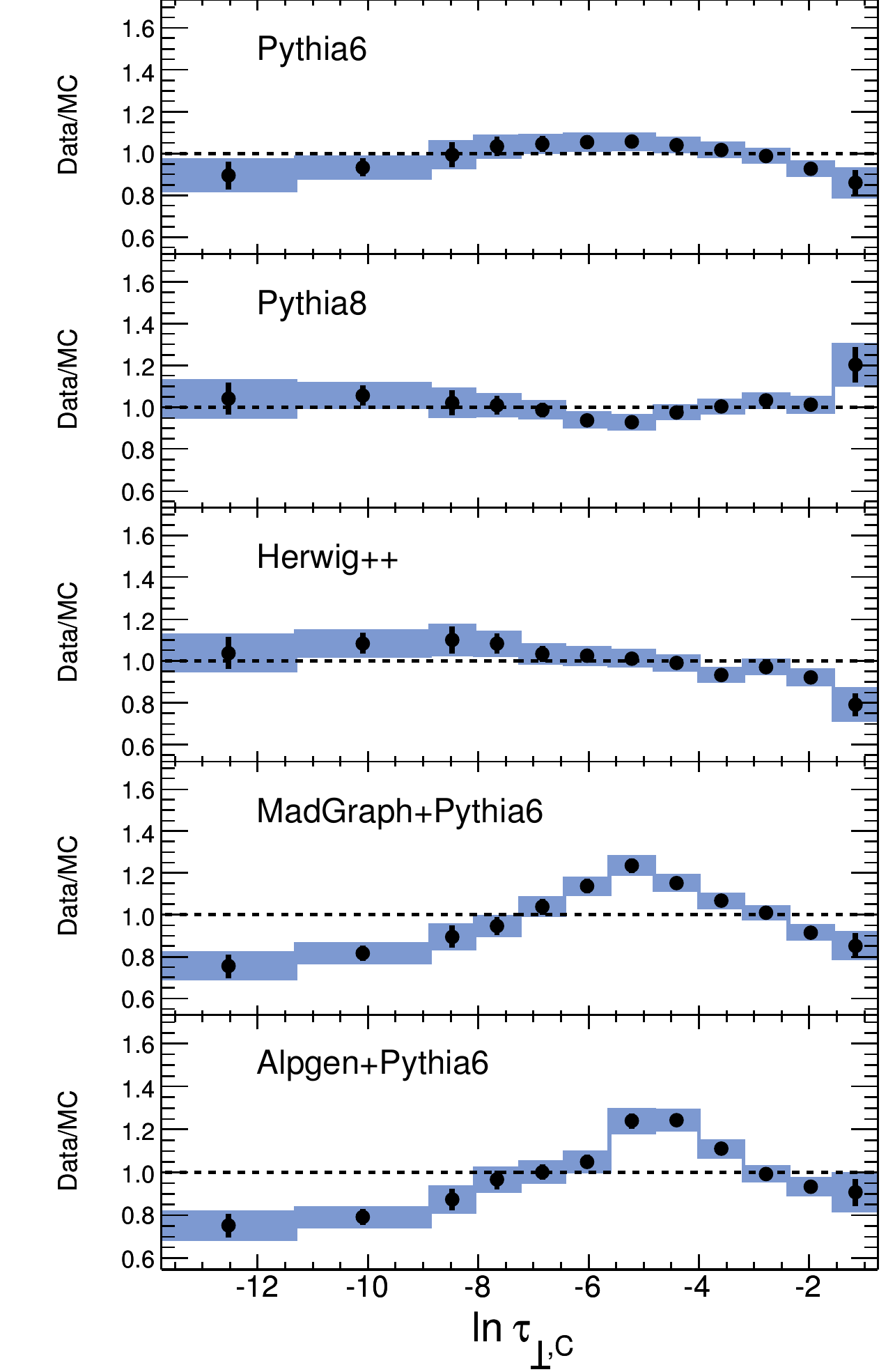}
    \includegraphics[width=0.49\textwidth]{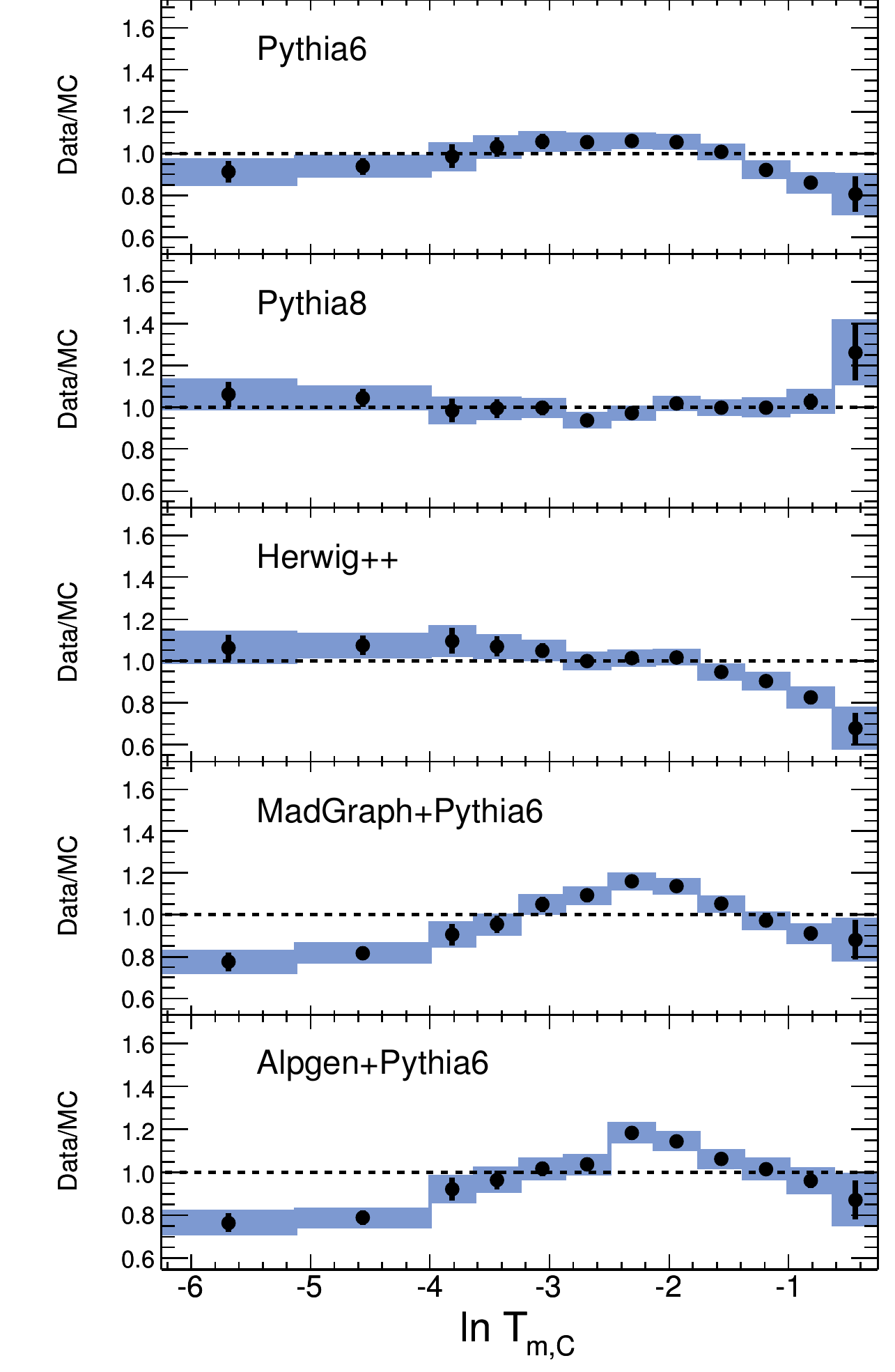}
    \caption{Distributions of the logarithm of the central transverse thrust (top left) and central thrust minor (top right)
             for events with a leading jet \highbin, from data and from five MC simulations.     				
    				   The error bars on the data points represent  the statistical uncertainty on the data, and
				     the shaded (blue) bands represent the sum of statistical and systematic errors.
             The lower plots show the ratio between data and the different simulated samples.}
    \label{fig:comparison3}
  \end{center}
\end{figure*}

The unfolded event-shape distributions from data and from the
\generators\ MC generators are shown in Figs.~\ref{fig:comparison1}--\ref{fig:comparison3} for the low, medium, and high-\pt samples, respectively.
The error bars represent the statistical uncertainties on the data and the shaded (blue) bands represent
the quadratic sum of the statistical and systematic uncertainties, discussed above.
The ratios between data and MC simulations are shown in the lower plots of Figs.~\ref{fig:comparison1}--\ref{fig:comparison3}
for each of the five MC generators.

The \pythiasix\ and \herwigpp\ predictions agree with the measurements in all three momentum bins,
while the \ALPGEN and \MADGRAPH curves deviate from the data as a result of an overestimate of the
fraction of back-to-back dijet events, which enter the lower tail of the distributions.
The \pythiaeight\ predictions agree with the measurements in all bins of the event-shape variables
except in the highest bin, where an underestimate is observed. This disagreement, however, affects only
a very small region of the parameter space (0.5\% of all events).

Further studies indicate that, while the momentum of the first leading jet agrees well between data and MC generators,
that of the second leading jet is higher in \MADGRAPH than in the data. This results in differences in the distribution of $\Delta\phi$
between the two leading jets, which then lead to a shift of the event-shape distributions towards lower values~\cite{DPHI}.
We conclude that the regime at high jet momentum and large number of jets, where the explicit ME calculations of
\ALPGEN and \MADGRAPH significantly improve the pure PS treatment, has not been reached.

In conclusion, we have presented the first measurement of two event-shape variables, the central transverse thrust
and the central thrust minor, using a data sample of proton-proton collisions at a center-of-mass energy of 7\TeV,
accumulated by the CMS detector at the LHC.
The measured event-shape distributions are presented after correction for the detector response.
We compare them with predictions from the \generators\ MC generators.
The event-shape distributions from \pythiasix, \pythiaeight, and \herwigpp\ show satisfactory agreement with the data, while
discrepancies are found between the data and predictions from \ALPGEN and \MADGRAPH.
These measurements provide input for the improvement of currently available models of QCD multijet production.

We wish to congratulate our colleagues in the CERN accelerator departments for the excellent performance of the LHC machine. We thank the technical and administrative staff at CERN and other CMS institutes, and acknowledge support from: FMSR (Austria); FNRS and FWO (Belgium); CNPq, CAPES, FAPERJ, and FAPESP (Brazil); MES (Bulgaria); CERN; CAS, MoST, and NSFC (China); COLCIENCIAS (Colombia); MSES (Croatia); RPF (Cyprus); Academy of Sciences and NICPB (Estonia); Academy of Finland, ME, and HIP (Finland); CEA and CNRS/IN2P3 (France); BMBF, DFG, and HGF (Germany); GSRT (Greece); OTKA and NKTH (Hungary); DAE and DST (India); IPM (Iran); SFI (Ireland); INFN (Italy); NRF and WCU (Korea); LAS (Lithuania); CINVESTAV, CONACYT, SEP, and UASLP-FAI (Mexico); PAEC (Pakistan); SCSR (Poland); FCT (Portugal); JINR (Armenia, Belarus, Georgia, Ukraine, Uzbekistan); MST and MAE (Russia); MSTD (Serbia); MICINN and CPAN (Spain); Swiss Funding Agencies (Switzerland); NSC (Taipei); TUBITAK and TAEK (Turkey); STFC (United Kingdom); DOE and NSF (USA). Individuals have received support from the Marie-Curie programme and the European Research Council (European Union); the Leventis Foundation; the A. P. Sloan Foundation; the Alexander von Humboldt Foundation; the Associazione per lo Sviluppo Scientifico e Tecnologico del Piemonte (Italy); the Belgian Federal Science Policy Office; the Fonds pour la Formation \`a la Recherche dans l'\'industrie et dans l'\'Agriculture (FRIA-Belgium); and the Agentschap voor Innovatie door Wetenschap en Technologie (IWT-Belgium).

\bibliography{auto_generated}   
\cleardoublepage\appendix\section{The CMS Collaboration \label{app:collab}}\begin{sloppypar}\hyphenpenalty=5000\widowpenalty=500\clubpenalty=5000\input{QCD-10-013-authorlist.tex}\end{sloppypar}
\end{document}

%% file: ptdr-definitions.tex
%
%
%

\providecommand {\etal}{\mbox{et al.}\xspace} 
\providecommand {\ie}{\mbox{i.e.}\xspace}     
\providecommand {\eg}{\mbox{e.g.}\xspace}     
\providecommand {\etc}{\mbox{etc.}\xspace}     
\providecommand {\vs}{\mbox{\sl vs.}\xspace}      
\providecommand {\mdash}{\ensuremath{\mathrm{-}}} 

\providecommand {\Lone}{Level-1\xspace} 
\providecommand {\Ltwo}{Level-2\xspace}
\providecommand {\Lthree}{Level-3\xspace}

\providecommand{\ACERMC} {\textsc{AcerMC}\xspace}
\providecommand{\ALPGEN} {{\textsc{alpgen}}\xspace}
\providecommand{\CHARYBDIS} {{\textsc{charybdis}}\xspace}
\providecommand{\CMKIN} {\textsc{cmkin}\xspace}
\providecommand{\CMSIM} {{\textsc{cmsim}}\xspace}
\providecommand{\CMSSW} {{\textsc{cmssw}}\xspace}
\providecommand{\COBRA} {{\textsc{cobra}}\xspace}
\providecommand{\COCOA} {{\textsc{cocoa}}\xspace}
\providecommand{\COMPHEP} {\textsc{CompHEP}\xspace}
\providecommand{\EVTGEN} {{\textsc{evtgen}}\xspace}
\providecommand{\FAMOS} {{\textsc{famos}}\xspace}
\providecommand{\GARCON} {\textsc{garcon}\xspace}
\providecommand{\GARFIELD} {{\textsc{garfield}}\xspace}
\providecommand{\GEANE} {{\textsc{geane}}\xspace}
\providecommand{\GEANTfour} {{\textsc{geant4}}\xspace}
\providecommand{\GEANTthree} {{\textsc{geant3}}\xspace}
\providecommand{\GEANT} {{\textsc{geant}}\xspace}
\providecommand{\HDECAY} {\textsc{hdecay}\xspace}
\providecommand{\HERWIG} {{\textsc{herwig}}\xspace}
\providecommand{\HIGLU} {{\textsc{higlu}}\xspace}
\providecommand{\HIJING} {{\textsc{hijing}}\xspace}
\providecommand{\IGUANA} {\textsc{iguana}\xspace}
\providecommand{\ISAJET} {{\textsc{isajet}}\xspace}
\providecommand{\ISAPYTHIA} {{\textsc{isapythia}}\xspace}
\providecommand{\ISASUGRA} {{\textsc{isasugra}}\xspace}
\providecommand{\ISASUSY} {{\textsc{isasusy}}\xspace}
\providecommand{\ISAWIG} {{\textsc{isawig}}\xspace}
\providecommand{\MADGRAPH} {\textsc{MadGraph}\xspace}
\providecommand{\MCATNLO} {\textsc{mc@nlo}\xspace}
\providecommand{\MCFM} {\textsc{mcfm}\xspace}
\providecommand{\MILLEPEDE} {{\textsc{millepede}}\xspace}
\providecommand{\ORCA} {{\textsc{orca}}\xspace}
\providecommand{\OSCAR} {{\textsc{oscar}}\xspace}
\providecommand{\PHOTOS} {\textsc{photos}\xspace}
\providecommand{\PROSPINO} {\textsc{prospino}\xspace}
\providecommand{\PYTHIA} {{\textsc{pythia}}\xspace}
\providecommand{\SHERPA} {{\textsc{sherpa}}\xspace}
\providecommand{\TAUOLA} {\textsc{tauola}\xspace}
\providecommand{\TOPREX} {\textsc{TopReX}\xspace}
\providecommand{\XDAQ} {{\textsc{xdaq}}\xspace}

\providecommand {\DZERO}{D\O\xspace}     


\providecommand{\de}{\ensuremath{^\circ}}
\providecommand{\ten}[1]{\ensuremath{\times \text{10}^\text{#1}}}
\providecommand{\unit}[1]{\ensuremath{\text{\,#1}}\xspace}
\providecommand{\mum}{\ensuremath{\,\mu\text{m}}\xspace}
\providecommand{\micron}{\ensuremath{\,\mu\text{m}}\xspace}
\providecommand{\cm}{\ensuremath{\,\text{cm}}\xspace}
\providecommand{\mm}{\ensuremath{\,\text{mm}}\xspace}
\providecommand{\mus}{\ensuremath{\,\mu\text{s}}\xspace}
\providecommand{\keV}{\ensuremath{\,\text{ke\hspace{-.08em}V}}\xspace}
\providecommand{\MeV}{\ensuremath{\,\text{Me\hspace{-.08em}V}}\xspace}
\providecommand{\GeV}{\ensuremath{\,\text{Ge\hspace{-.08em}V}}\xspace}
\providecommand{\gev}{\GeV}
\providecommand{\TeV}{\ensuremath{\,\text{Te\hspace{-.08em}V}}\xspace}
\providecommand{\PeV}{\ensuremath{\,\text{Pe\hspace{-.08em}V}}\xspace}
\providecommand{\keVc}{\ensuremath{{\,\text{ke\hspace{-.08em}V\hspace{-0.16em}/\hspace{-0.08em}}c}}\xspace}
\providecommand{\MeVc}{\ensuremath{{\,\text{Me\hspace{-.08em}V\hspace{-0.16em}/\hspace{-0.08em}}c}}\xspace}
\providecommand{\GeVc}{\ensuremath{{\,\text{Ge\hspace{-.08em}V\hspace{-0.16em}/\hspace{-0.08em}}c}}\xspace}
\providecommand{\TeVc}{\ensuremath{{\,\text{Te\hspace{-.08em}V\hspace{-0.16em}/\hspace{-0.08em}}c}}\xspace}
\providecommand{\keVcc}{\ensuremath{{\,\text{ke\hspace{-.08em}V\hspace{-0.16em}/\hspace{-0.08em}}c^\text{2}}}\xspace}
\providecommand{\MeVcc}{\ensuremath{{\,\text{Me\hspace{-.08em}V\hspace{-0.16em}/\hspace{-0.08em}}c^\text{2}}}\xspace}
\providecommand{\GeVcc}{\ensuremath{{\,\text{Ge\hspace{-.08em}V\hspace{-0.16em}/\hspace{-0.08em}}c^\text{2}}}\xspace}
\providecommand{\TeVcc}{\ensuremath{{\,\text{Te\hspace{-.08em}V\hspace{-0.16em}/\hspace{-0.08em}}c^\text{2}}}\xspace}

\providecommand{\pbinv} {\mbox{\ensuremath{\,\text{pb}^\text{$-$1}}}\xspace}
\providecommand{\fbinv} {\mbox{\ensuremath{\,\text{fb}^\text{$-$1}}}\xspace}
\providecommand{\nbinv} {\mbox{\ensuremath{\,\text{nb}^\text{$-$1}}}\xspace}
\providecommand{\percms}{\ensuremath{\,\text{cm}^\text{$-$2}\,\text{s}^\text{$-$1}}\xspace}
\providecommand{\lumi}{\ensuremath{\mathcal{L}}\xspace}
\providecommand{\Lumi}{\ensuremath{\mathcal{L}}\xspace}
%
\providecommand{\LvLow}  {\ensuremath{\mathcal{L}=\text{10}^\text{32}\,\text{cm}^\text{$-$2}\,\text{s}^\text{$-$1}}\xspace}
\providecommand{\LLow}   {\ensuremath{\mathcal{L}=\text{10}^\text{33}\,\text{cm}^\text{$-$2}\,\text{s}^\text{$-$1}}\xspace}
\providecommand{\lowlumi}{\ensuremath{\mathcal{L}=\text{2}\times \text{10}^\text{33}\,\text{cm}^\text{$-$2}\,\text{s}^\text{$-$1}}\xspace}
\providecommand{\LMed}   {\ensuremath{\mathcal{L}=\text{2}\times \text{10}^\text{33}\,\text{cm}^\text{$-$2}\,\text{s}^\text{$-$1}}\xspace}
\providecommand{\LHigh}  {\ensuremath{\mathcal{L}=\text{10}^\text{34}\,\text{cm}^\text{$-$2}\,\text{s}^\text{$-$1}}\xspace}
\providecommand{\hilumi} {\ensuremath{\mathcal{L}=\text{10}^\text{34}\,\text{cm}^\text{$-$2}\,\text{s}^\text{$-$1}}\xspace}


\providecommand{\PT}{\ensuremath{p_{\mathrm{T}}}\xspace}
\providecommand{\pt}{\ensuremath{p_{\mathrm{T}}}\xspace}
\providecommand{\ET}{\ensuremath{E_{\mathrm{T}}}\xspace}
\providecommand{\HT}{\ensuremath{H_{\mathrm{T}}}\xspace}
\providecommand{\et}{\ensuremath{E_{\mathrm{T}}}\xspace}
\providecommand{\Em}{\ensuremath{E\hspace{-0.6em}/}\xspace}
\providecommand{\Pm}{\ensuremath{p\hspace{-0.5em}/}\xspace}
\providecommand{\PTm}{\ensuremath{{p}_\mathrm{T}\hspace{-1.02em}/}\xspace}
\providecommand{\PTslash}{\ensuremath{{p}_\mathrm{T}\hspace{-1.02em}/}\xspace}
\providecommand{\ETm}{\ensuremath{E_{\mathrm{T}}^{\text{miss}}}\xspace}
\providecommand{\MET}{\ETm}
\providecommand{\ETmiss}{\ETm}
\providecommand{\ETslash}{\ensuremath{E_{\mathrm{T}}\hspace{-1.1em}/}\xspace}
\providecommand{\VEtmiss}{\ensuremath{{\vec E}_{\mathrm{T}}^{\text{miss}}}\xspace}

\providecommand{\dd}[2]{\ensuremath{\frac{\mathrm{d} #1}{\mathrm{d} #2}}}
\providecommand{\ddinline}[2]{\ensuremath{\mathrm{d} #1/\mathrm{d} #2}}

\ifthenelse{\boolean{cms@italic}}{\newcommand{\cmsSymbolFace}{\relax}}{\newcommand{\cmsSymbolFace}{\mathrm}}

\providecommand{\zp}{\ensuremath{\cmsSymbolFace{Z}^\prime}\xspace}
\providecommand{\JPsi}{\ensuremath{\cmsSymbolFace{J}\hspace{-.08em}/\hspace{-.14em}\psi}\xspace}
\providecommand{\Z}{\ensuremath{\cmsSymbolFace{Z}}\xspace}
\providecommand{\ttbar}{\ensuremath{\cmsSymbolFace{t}\overline{\cmsSymbolFace{t}}}\xspace}

\newcommand{\cPgn}{\ensuremath{\nu}}
\newcommand{\cPJgy}{\JPsi}
\newcommand{\cPZ}{\Z}
\newcommand{\cPZpr}{\zp}


\providecommand{\AFB}{\ensuremath{A_\text{FB}}\xspace}
\providecommand{\wangle}{\ensuremath{\sin^{2}\theta_{\text{eff}}^\text{lept}(M^2_\Z)}\xspace}
\providecommand{\stat}{\ensuremath{\,\text{(stat.)}}\xspace}
\providecommand{\syst}{\ensuremath{\,\text{(syst.)}}\xspace}
\providecommand{\kt}{\ensuremath{k_{\mathrm{T}}}\xspace}

\providecommand{\BC}{\ensuremath{\mathrm{B_{c}}}\xspace}
\providecommand{\bbarc}{\ensuremath{\mathrm{\overline{b}c}}\xspace}
\providecommand{\bbbar}{\ensuremath{\mathrm{b\overline{b}}}\xspace}
\providecommand{\ccbar}{\ensuremath{\mathrm{c\overline{c}}}\xspace}
\providecommand{\bspsiphi}{\ensuremath{\mathrm{B_s} \to \JPsi\, \phi}\xspace}
\providecommand{\EE}{\ensuremath{\mathrm{e^+e^-}}\xspace}
\providecommand{\MM}{\ensuremath{\mu^+\mu^-}\xspace}
\providecommand{\TT}{\ensuremath{\tau^+\tau^-}\xspace}

\providecommand{\HGG}{\ensuremath{\mathrm{H}\to\gamma\gamma}}
\providecommand{\GAMJET}{\ensuremath{\gamma + \text{jet}}}
\providecommand{\PPTOJETS}{\ensuremath{\mathrm{pp}\to\text{jets}}}
\providecommand{\PPTOGG}{\ensuremath{\mathrm{pp}\to\gamma\gamma}}
\providecommand{\PPTOGAMJET}{\ensuremath{\mathrm{pp}\to\gamma + \mathrm{jet}}}
\providecommand{\MH}{\ensuremath{M_{\mathrm{H}}}}
\providecommand{\RNINE}{\ensuremath{R_\mathrm{9}}}
\providecommand{\DR}{\ensuremath{\Delta R}}

%

\providecommand{\ga}{\ensuremath{\gtrsim}}
\providecommand{\la}{\ensuremath{\lesssim}}
\providecommand{\swsq}{\ensuremath{\sin^2\theta_\cmsSymbolFace{W}}\xspace}
\providecommand{\cwsq}{\ensuremath{\cos^2\theta_\cmsSymbolFace{W}}\xspace}
\providecommand{\tanb}{\ensuremath{\tan\beta}\xspace}
\providecommand{\tanbsq}{\ensuremath{\tan^{2}\beta}\xspace}
\providecommand{\sidb}{\ensuremath{\sin 2\beta}\xspace}
\providecommand{\alpS}{\ensuremath{\alpha_S}\xspace}
\providecommand{\alpt}{\ensuremath{\tilde{\alpha}}\xspace}

\providecommand{\QL}{\ensuremath{\cmsSymbolFace{Q}_\cmsSymbolFace{L}}\xspace}
\providecommand{\sQ}{\ensuremath{\tilde{\cmsSymbolFace{Q}}}\xspace}
\providecommand{\sQL}{\ensuremath{\tilde{\cmsSymbolFace{Q}}_\cmsSymbolFace{L}}\xspace}
\providecommand{\ULC}{\ensuremath{\cmsSymbolFace{U}_\cmsSymbolFace{L}^\cmsSymbolFace{C}}\xspace}
\providecommand{\sUC}{\ensuremath{\tilde{\cmsSymbolFace{U}}^\cmsSymbolFace{C}}\xspace}
\providecommand{\sULC}{\ensuremath{\tilde{\cmsSymbolFace{U}}_\cmsSymbolFace{L}^\cmsSymbolFace{C}}\xspace}
\providecommand{\DLC}{\ensuremath{\cmsSymbolFace{D}_\cmsSymbolFace{L}^\cmsSymbolFace{C}}\xspace}
\providecommand{\sDC}{\ensuremath{\tilde{\cmsSymbolFace{D}}^\cmsSymbolFace{C}}\xspace}
\providecommand{\sDLC}{\ensuremath{\tilde{\cmsSymbolFace{D}}_\cmsSymbolFace{L}^\cmsSymbolFace{C}}\xspace}
\providecommand{\LL}{\ensuremath{\cmsSymbolFace{L}_\cmsSymbolFace{L}}\xspace}
\providecommand{\sL}{\ensuremath{\tilde{\cmsSymbolFace{L}}}\xspace}
\providecommand{\sLL}{\ensuremath{\tilde{\cmsSymbolFace{L}}_\cmsSymbolFace{L}}\xspace}
\providecommand{\ELC}{\ensuremath{\cmsSymbolFace{E}_\cmsSymbolFace{L}^\cmsSymbolFace{C}}\xspace}
\providecommand{\sEC}{\ensuremath{\tilde{\cmsSymbolFace{E}}^\cmsSymbolFace{C}}\xspace}
\providecommand{\sELC}{\ensuremath{\tilde{\cmsSymbolFace{E}}_\cmsSymbolFace{L}^\cmsSymbolFace{C}}\xspace}
\providecommand{\sEL}{\ensuremath{\tilde{\cmsSymbolFace{E}}_\cmsSymbolFace{L}}\xspace}
\providecommand{\sER}{\ensuremath{\tilde{\cmsSymbolFace{E}}_\cmsSymbolFace{R}}\xspace}
\providecommand{\sFer}{\ensuremath{\tilde{\cmsSymbolFace{f}}}\xspace}
\providecommand{\sQua}{\ensuremath{\tilde{\cmsSymbolFace{q}}}\xspace}
\providecommand{\sUp}{\ensuremath{\tilde{\cmsSymbolFace{u}}}\xspace}
\providecommand{\suL}{\ensuremath{\tilde{\cmsSymbolFace{u}}_\cmsSymbolFace{L}}\xspace}
\providecommand{\suR}{\ensuremath{\tilde{\cmsSymbolFace{u}}_\cmsSymbolFace{R}}\xspace}
\providecommand{\sDw}{\ensuremath{\tilde{\cmsSymbolFace{d}}}\xspace}
\providecommand{\sdL}{\ensuremath{\tilde{\cmsSymbolFace{d}}_\cmsSymbolFace{L}}\xspace}
\providecommand{\sdR}{\ensuremath{\tilde{\cmsSymbolFace{d}}_\cmsSymbolFace{R}}\xspace}
\providecommand{\sTop}{\ensuremath{\tilde{\cmsSymbolFace{t}}}\xspace}
\providecommand{\stL}{\ensuremath{\tilde{\cmsSymbolFace{t}}_\cmsSymbolFace{L}}\xspace}
\providecommand{\stR}{\ensuremath{\tilde{\cmsSymbolFace{t}}_\cmsSymbolFace{R}}\xspace}
\providecommand{\stone}{\ensuremath{\tilde{\cmsSymbolFace{t}}_1}\xspace}
\providecommand{\sttwo}{\ensuremath{\tilde{\cmsSymbolFace{t}}_2}\xspace}
\providecommand{\sBot}{\ensuremath{\tilde{\cmsSymbolFace{b}}}\xspace}
\providecommand{\sbL}{\ensuremath{\tilde{\cmsSymbolFace{b}}_\cmsSymbolFace{L}}\xspace}
\providecommand{\sbR}{\ensuremath{\tilde{\cmsSymbolFace{b}}_\cmsSymbolFace{R}}\xspace}
\providecommand{\sbone}{\ensuremath{\tilde{\cmsSymbolFace{b}}_1}\xspace}
\providecommand{\sbtwo}{\ensuremath{\tilde{\cmsSymbolFace{b}}_2}\xspace}
\providecommand{\sLep}{\ensuremath{\tilde{\cmsSymbolFace{l}}}\xspace}
\providecommand{\sLepC}{\ensuremath{\tilde{\cmsSymbolFace{l}}^\cmsSymbolFace{C}}\xspace}
\providecommand{\sEl}{\ensuremath{\tilde{\cmsSymbolFace{e}}}\xspace}
\providecommand{\sElC}{\ensuremath{\tilde{\cmsSymbolFace{e}}^\cmsSymbolFace{C}}\xspace}
\providecommand{\seL}{\ensuremath{\tilde{\cmsSymbolFace{e}}_\cmsSymbolFace{L}}\xspace}
\providecommand{\seR}{\ensuremath{\tilde{\cmsSymbolFace{e}}_\cmsSymbolFace{R}}\xspace}
\providecommand{\snL}{\ensuremath{\tilde{\nu}_L}\xspace}
\providecommand{\sMu}{\ensuremath{\tilde{\mu}}\xspace}
\providecommand{\sNu}{\ensuremath{\tilde{\nu}}\xspace}
\providecommand{\sTau}{\ensuremath{\tilde{\tau}}\xspace}
\providecommand{\Glu}{\ensuremath{\cmsSymbolFace{g}}\xspace}
\providecommand{\sGlu}{\ensuremath{\tilde{\cmsSymbolFace{g}}}\xspace}
\providecommand{\Wpm}{\ensuremath{\cmsSymbolFace{W}^{\pm}}\xspace}
\providecommand{\sWpm}{\ensuremath{\tilde{\cmsSymbolFace{W}}^{\pm}}\xspace}
\providecommand{\Wz}{\ensuremath{\cmsSymbolFace{W}^{0}}\xspace}
\providecommand{\sWz}{\ensuremath{\tilde{\cmsSymbolFace{W}}^{0}}\xspace}
\providecommand{\sWino}{\ensuremath{\tilde{\cmsSymbolFace{W}}}\xspace}
\providecommand{\Bz}{\ensuremath{\cmsSymbolFace{B}^{0}}\xspace}
\providecommand{\sBz}{\ensuremath{\tilde{\cmsSymbolFace{B}}^{0}}\xspace}
\providecommand{\sBino}{\ensuremath{\tilde{\cmsSymbolFace{B}}}\xspace}
\providecommand{\Zz}{\ensuremath{\cmsSymbolFace{Z}^{0}}\xspace}
\providecommand{\sZino}{\ensuremath{\tilde{\cmsSymbolFace{Z}}^{0}}\xspace}
\providecommand{\sGam}{\ensuremath{\tilde{\gamma}}\xspace}
\providecommand{\chiz}{\ensuremath{\tilde{\chi}^{0}}\xspace}
\providecommand{\chip}{\ensuremath{\tilde{\chi}^{+}}\xspace}
\providecommand{\chim}{\ensuremath{\tilde{\chi}^{-}}\xspace}
\providecommand{\chipm}{\ensuremath{\tilde{\chi}^{\pm}}\xspace}
\providecommand{\Hone}{\ensuremath{\cmsSymbolFace{H}_\cmsSymbolFace{d}}\xspace}
\providecommand{\sHone}{\ensuremath{\tilde{\cmsSymbolFace{H}}_\cmsSymbolFace{d}}\xspace}
\providecommand{\Htwo}{\ensuremath{\cmsSymbolFace{H}_\cmsSymbolFace{u}}\xspace}
\providecommand{\sHtwo}{\ensuremath{\tilde{\cmsSymbolFace{H}}_\cmsSymbolFace{u}}\xspace}
\providecommand{\sHig}{\ensuremath{\tilde{\cmsSymbolFace{H}}}\xspace}
\providecommand{\sHa}{\ensuremath{\tilde{\cmsSymbolFace{H}}_\cmsSymbolFace{a}}\xspace}
\providecommand{\sHb}{\ensuremath{\tilde{\cmsSymbolFace{H}}_\cmsSymbolFace{b}}\xspace}
\providecommand{\sHpm}{\ensuremath{\tilde{\cmsSymbolFace{H}}^{\pm}}\xspace}
\providecommand{\hz}{\ensuremath{\cmsSymbolFace{h}^{0}}\xspace}
\providecommand{\Hz}{\ensuremath{\cmsSymbolFace{H}^{0}}\xspace}
\providecommand{\Az}{\ensuremath{\cmsSymbolFace{A}^{0}}\xspace}
\providecommand{\Hpm}{\ensuremath{\cmsSymbolFace{H}^{\pm}}\xspace}
\providecommand{\sGra}{\ensuremath{\tilde{\cmsSymbolFace{G}}}\xspace}
\providecommand{\mtil}{\ensuremath{\tilde{m}}\xspace}
\providecommand{\rpv}{\ensuremath{\rlap{\kern.2em/}R}\xspace}
\providecommand{\LLE}{\ensuremath{LL\bar{E}}\xspace}
\providecommand{\LQD}{\ensuremath{LQ\bar{D}}\xspace}
\providecommand{\UDD}{\ensuremath{\overline{UDD}}\xspace}
\providecommand{\Lam}{\ensuremath{\lambda}\xspace}
\providecommand{\Lamp}{\ensuremath{\lambda'}\xspace}
\providecommand{\Lampp}{\ensuremath{\lambda''}\xspace}
\providecommand{\spinbd}[2]{\ensuremath{\bar{#1}_{\dot{#2}}}\xspace}

\providecommand{\MD}{\ensuremath{{M_\mathrm{D}}}\xspace}
\providecommand{\Mpl}{\ensuremath{{M_\mathrm{Pl}}}\xspace}
\providecommand{\Rinv} {\ensuremath{{R}^{-1}}\xspace} 

%% file: QCD-10-013-authorlist.tex
\textbf{Yerevan Physics Institute,  Yerevan,  Armenia}\\*[0pt]
V.~Khachatryan, A.M.~Sirunyan, A.~Tumasyan
\vskip\cmsinstskip
\textbf{Institut f\"{u}r Hochenergiephysik der OeAW,  Wien,  Austria}\\*[0pt]
W.~Adam, T.~Bergauer, M.~Dragicevic, J.~Er\"{o}, C.~Fabjan, M.~Friedl, R.~Fr\"{u}hwirth, V.M.~Ghete, J.~Hammer\cmsAuthorMark{1}, S.~H\"{a}nsel, C.~Hartl, M.~Hoch, N.~H\"{o}rmann, J.~Hrubec, M.~Jeitler, G.~Kasieczka, W.~Kiesenhofer, M.~Krammer, D.~Liko, I.~Mikulec, M.~Pernicka, H.~Rohringer, R.~Sch\"{o}fbeck, J.~Strauss, A.~Taurok, F.~Teischinger, W.~Waltenberger, G.~Walzel, E.~Widl, C.-E.~Wulz
% ...
\vskip\cmsinstskip
% ...
1:~~Also at CERN, European Organization for Nuclear Research, Geneva, Switzerland\\
% ...
